\documentclass[aps,pra,showpacs,groupedaddress,twocolumn,preprintnumbers,amsmath,amssymb]{revtex4}
\usepackage{graphicx}
\usepackage{color}
\usepackage{dcolumn}
\usepackage{bm}
\usepackage{amsmath}

\begin{document}

\title{Single Nuclear Spin Cavity QED}

\author{Makoto Takeuchi$^{1,\dag }$, Nobuyuki Takei$^{1,\ddag }$, Kodai Doi$^{1,3}$, Peng Zhang$^{1}$, Masahito Ueda$^{1,2}$, and Mikio Kozuma$^{1,3}$}

\affiliation{%
$^{1}$ERATO Macroscopic Quantum Control Project, JST, 2-11-16 Yayoi, Bunkyo-Ku, Tokyo 113-8656, Japan}

\affiliation{%
$^{2}$Department of Physics, University of Tokyo, Hongo, Bunkyo-ku, Tokyo 113-0033, Japan}

\affiliation{%
$^{3}$Department of Physics, Tokyo Institute of Technology, 2-12-1 O-okayama, Meguro-ku, Tokyo 152-8550, Japan}

\altaffiliation[Present affiliation:]{$\dag$ National Institute of Information and Communications Technology,
$\ddag$ Institute for Molecular Science}

\date{\today}

\begin{abstract}
We constructed a cavity QED system with a diamagnetic atom of $\mathrm{^{171}Yb}$ and performed projective measurements on a single nuclear spin.
Since Yb has no electronic spin and has 1/2 nuclear spin, the procedure of spin polarization and state verification can be dramatically simplified compared with the pseudo spin-1/2 system.
By enhancing the photon emission rate of the $^1S_0-{}^3P_1$ transition, projective measurement is implemented for an atom with the measurement time of $T_\mathrm{meas}=30~\mathrm{\mu s}$. Unwanted spin flip as well as dark counts of the detector lead to systematic error when the present technique is applied for the determination of diagonal elements of an unknown spin state, which is $\delta|\beta|^2\le2\times 10^{-2}$.
Fast measurement on a long-lived qubit is key to the realization of large-scale one-way quantum computing.
\end{abstract}

\maketitle

\section{introduction}

Quantum information processing with neutral-atom qubits is advantageous when a large number of qubits are required.
The most prominent advantage is easy production of the large-scale cluster state \cite{Mandel03},
which is a resource state for one-way quantum computing \cite{Raussendorf01}.
One-way quantum computing consists of the following four stages: (i) preparation of an optical lattice filled by single atoms; (ii) creation of cluster state among them; (iii) loading them into measurement region by moving lattice \cite{Fortier07}; (iv) one-by-one projective measurements and feedforward on a part of them. 
In the last stage, we must keep the available number of measurements and feedforward within the coherence time. 
Therefore fast measurement on a long-lived qubit is key to the realization of large-scale quantum computing. As a long-lived qubit, a nuclear spin in a diamagnetic atom is promising.
A diamagnetic atom has a small magnetic moment, which is three orders of magnitude smaller than that of a paramagnetic atom, originating from its nuclear spin. 
Accordingly, the decoherence caused by stray magnetic fields can be greatly suppressed. 
As a fast measurement technique, enhanced spontaneous emission (ESE) which is utilizable in cavity quantum electrodynamics (QED) systems is helpful \cite{Kleppner81,Heinzen87-1, Vahala03}.
ESE is a phenomenon whereby an atom in a cavity emits photons into the output mode of the cavity faster than into free space. 
Since cavity QED systems have primarily been realized with paramagnetic atoms, 
one might think that clock states are also promising \cite{Kreuter04, Weber09, Boozer07, Russo09}.
However, other extra substates in its ground state interrupt directly observing the clock states. Note that cavity QED experiments performed so far have addressed not clock states but hyperfine substates \cite{Boozer06, Khudaverdyan09}.
Mapping a clock state population to a cycling transitions, demonstrated with paramagnetic ion, is inapplicable to conventional cavity QED systems because nuclear spin $I$ of atoms utilized in these experiments are not 1/2 \cite{Acton06}.

Here, we report the construction of a cavity QED system with a diamagnetic atom of $\mathrm{^{171}Yb}$ and the observation of a single nuclear spin.
We have enhanced the photon emission rate of the $^1S_0-{}^3P_1$ transition, where the resonant wavelength is $\lambda=556~$nm, and the natural linewidth is $\gamma=2\pi\times 182~$kHz.
The cavity is of the Fabry-Perot type, where the maximum coupling strength between an atom and the cavity is $g_\mathrm{max}=2\pi\times2.8$~MHz.
Since the optical lattice filled with single Yb atoms has been recently demonstrated \cite{Fukuhara09} and the method to produce the cluster state for diamagnetic atoms are proposed \cite{Daley08}, it should be possible to implement the nuclear-spin based quantum computing with $\mathrm{^{171}Yb}$ atoms.

\section{Experimental setup}
\begin{figure}[htbp]
\begin{center}
\scalebox{0.30}{\includegraphics{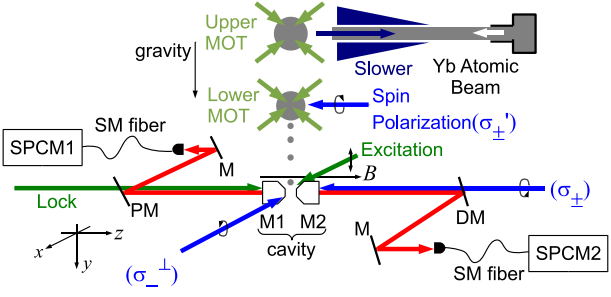}}
\caption{(Color online)
Overview of the experimental apparatus.
M: mirror, DM: dichroic mirror, PM: partially reflecting mirror.
Yb atoms are Zeeman slowed (399~nm) and are trapped in the upper magneto-optical trap (MOT, 556nm).
The atoms released from the upper MOT are recaptured by the lower MOT.
A light beam with a wavelength of 399~nm and a power of 100~mW is generated by second harmonic generation using a bow-tie cavity with a Ti:Sapphire laser operated at a power of 0.6~W. A light beam having a wavelength of 556~nm and a power of 50~mW is obtained by the same manner as a fiber laser with a power of 0.1~W. The frequencies of these lasers are locked to two respective ULE reference cavities.The excitation beam is injected into the space between the two mirrors M1 and M2. 
}
\label{setup5}
\end{center}
\end{figure}

A single $\mathrm{^{171}Yb}$ atom is loaded to the cavity mode by free falling from a lower MOT as shown in Fig. \ref{setup5}.
Its position is approximately $H=7$~mm above the cavity and the atoms reach the cavity after $T_\mathrm{fall}=40$~ms with the velocity of $v_f=0.3$~m/s.
The velocity distributions of atoms after released from MOT along the $z$-axis and $x$-axis are evaluated to be $v_a=4\times 10^{-2}$~m/s by the absorption imaging \cite{Kuwamoto99}.
The spin can be initialized by a circularly polarized pulse ($\sigma_\pm,\sigma_-^\perp,\sigma_\pm'$) while or before an atom transits the cavity mode, which is resonant with the $^1S_0-{}^1P_1(F'=1/2)$ transition (399~nm).
The cavity consists of two concave mirrors M1 and M2.
Each mirror is glued to a piezoelectric transducer (PZT) and the spacing of the mirrors is $L_c=150~\mathrm{\mu m}$.
The curvature radius, reflectivity, transmittance, and loss of the mirror are $R_c=50$~mm, $R_m=0.999972$, $T_m=2.5\times10^{-5}$, and $L_m=3\times10^{-6}$, respectively.
The beam waist of the cavity mode is $w_c=19~\mathrm{\mu m}$.
Note that the mirrors of which the cavity consists show relatively high transparency of 75\% at the wavelength of 399nm. Thus by simply injecting the laser beams into the cavity, we could accomplish polarizing the spin state of the atom in the cavity.
The resonant frequency of the cavity $\omega_c$ is locked by using a FM-sideband method with a locking beam.
While observing photons emitted from the cavity during $T_\mathrm{hold}=$3ms, the locking beam is turned off by the sample-and-hold method.
The cavity decay rate is $\kappa=2\pi\times 4.5$~MHz.
An excitation beam with frequency $\omega_l$, power $P_\mathrm{total}=0.9~\mathrm{\mu W}$, and beam waist $w_l=24~\mathrm{\mu m}$ is injected into the space between the mirrors ($x$-axis).
The polarization of the excitation beam is linear ($y$-axis) and can be decomposed
 into $\sigma_+$ and $\sigma_-$ components for the quantization axis ($z$-axis).
The Rabi frequency for the $\sigma_-$ component at the center of the Gaussian profile is
 $\Omega_\mathrm{max}=2\pi\times 2.4$~MHz.
The transit time of an atom passing through the cavity mode is typically $T_\mathrm{transit} = 120~\mathrm{\mu}$s.
In order to ensure that the atom-cavity coupling is constant,
 we observe the atom only when it is close to the mode axis within the time window of $T_\mathrm{win} = 36~\mathrm{\mu s}$.
The mean travel distance along the $z$-direction during $T_\mathrm{win}$
 can be estimated as $v_a T_\mathrm{win}=1~\mathrm{\mu m}$,
 which is five times as large as the period of the standing wave $\lambda/2$ ($\lambda=$556~nm).

\section{Method of single nuclear spin detection}
\begin{figure}[htbp]
\begin{center}
\scalebox{0.35}{\includegraphics{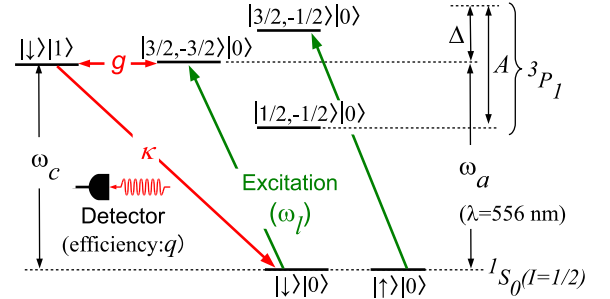}}
\caption{(Color online)
Energy-level diagram of $\mathrm{^{171}Yb}$ and the cavity used in the present experiment. 
The Zeeman substates $m_I=+1/2$ and $-1/2$ of the ground state $^1S_0(I=1/2)$
 are denoted by $|\!\!\uparrow\rangle$ and $|\!\!\downarrow\rangle$, respectively.
The substates in the excited state $^3P_1$ are labeled as $|F',m_{F'}\rangle$.
The number state of $n$ photons in the cavity mode is denoted by $|n\rangle$.
Each energy level is denoted as the product state of the atomic state and photon state $|n\rangle$.
Here, we have omitted unimportant excited states.
}
\label{energy}
\end{center}
\end{figure}

In Fig. \ref{energy}, we show the energy levels of $\mathrm{^{171}Yb}$
 which is coupled to the cavity.
The hyperfine splitting of the excited state is $A=2\pi\times 5.9~\mathrm{GHz}$.
The frequency $\omega_a$ represents the $|\!\!\downarrow\rangle\leftrightarrow|3/2,-3/2\rangle$ transition frequency.
A homogeneous magnetic field of $B=34$~Gauss is applied along the $z$ axis and the 
resultant Zeeman shift for the $|3/2,-3/2\rangle$ state is $\Delta=2\pi\times 71~$MHz.
Since the vacuum chamber is enclosed by $\mu$-metal,
 the residual magnetic field is suppressed to $2\times 10^{-2}$~Gauss.
Note that the coherence time is sensitive not to the static but to fluctuating magnetic field \cite{Langer05}.
In order to induce the ESE, the frequencies are tuned to $\omega_l=\omega_a=\omega_c$.
The atom in the $|\!\!\uparrow\rangle$ state seldom absorbs photons,
 because the excitation beam is far detuned from any transition.
In contrast, the atom in the $|\!\!\downarrow\rangle$ state can resonantly absorb a photon, and the population oscillates between the two levels
 $|3/2,-3/2\rangle|0\rangle$ and $|\!\!\downarrow\rangle|1\rangle$, due to the coupling $g$ between an atom and the cavity. After the oscillation, photons leak from the cavity at a rate of $\kappa$ and the atom decays back to the ground state $|\!\!\downarrow\rangle|0\rangle$. The emitted photons are coupled to single-mode fibers (SM fibers),
 and are detected by single photon counting modules
 (SPCM, PerkinElmer SPCM-AQR-14-FC).
These detection systems are placed on both sides of the output mode and the total detection efficiency for a photon emitted
 from an atom with the two SPCMs is $q=0.3$.

If the $|\!\!\uparrow\rangle$ atom is excited and decays to the state $|\!\!\downarrow\rangle$, photons will be repeatedly emitted from the spin-flipped atom, which limits the precision of this scheme.
Under the approximation of $g^2\gg \kappa\gamma$
 and $\Omega^2 \ll g^2\kappa/\gamma$, the intra-cavity photon number is given by
 $\langle n\rangle \sim\Omega^2/(4g^2)$, and the photon emission rate into
 the output mode of the cavity is $\Gamma=2\kappa \langle n\rangle\sim\kappa\Omega^2/(2g^2) $\cite{Kuhn99}. 
When a $|\!\!\downarrow\rangle$ atom locates
 at the crossed position of the cavity-mode axis
 and the center of the excitation beam profile,
 these values become maximal, i.e., 
$\langle n\rangle_\mathrm{max}=0.16$ and
$\Gamma_\mathrm{max}=9.3\times 10^6~\mathrm{s^{-1}}$.
As an effective value, we adopt $\Gamma=\Gamma_\mathrm{max}/2$ caused by motion of single atoms along the cavity axis.
The factor 1/2 is originated from the spatial dependence of $g^2$ as $\sin^2(2\pi z/\lambda)$. 
The unwanted spin-flip rate is roughly estimated to be $\Gamma_\mathrm{flip}=\Gamma^\mathrm{(free)}(\Delta)+\Gamma^\mathrm{(free)}(A-\Delta)$, where
\begin{align}
\Gamma^\mathrm{(free)}(\Delta)=\frac{\gamma}{2}\frac{\Omega^2/2}{\Delta^2+\gamma^2/4+\Omega^2/2} \label{Gamma^free}
\end{align}
is the photon absorption rate when an atom is located in free space illuminated by $\Delta$ detuned light.
In our experimental condition, $\Gamma_\mathrm{flip}=3\times 10^2~\mathrm{s^{-1}}$.
We define $S/N$ as the achievable number of photon counts from the $|\!\!\downarrow\rangle$ atom without the unwanted spin flip.
In our experiment, $S/N=q\Gamma/\Gamma_\mathrm{flip}=5\times 10^3$ is expected.
$S/N$ is maximized at $\Delta= A/2$, which gives $S/N\le q\kappa A^2/(2\gamma g^2)$.
Our method is applicable for a narrow transition linewidth which has a large hyperfine splitting in the excited state.
In the case of electric dipole transition usually used for cavity QED experiments,
$S/N$ usually becomes small. 
It is the order of $S/N\le 10$ even assuming the detection efficiency to be perfect and the applied magnetic field to be optimal because of small $A$ and large $\gamma$ \cite{Boozer06}.

\section{Observation of an atomic transit}
\begin{figure}[htbp]
\begin{center}
\scalebox{0.2}{\includegraphics{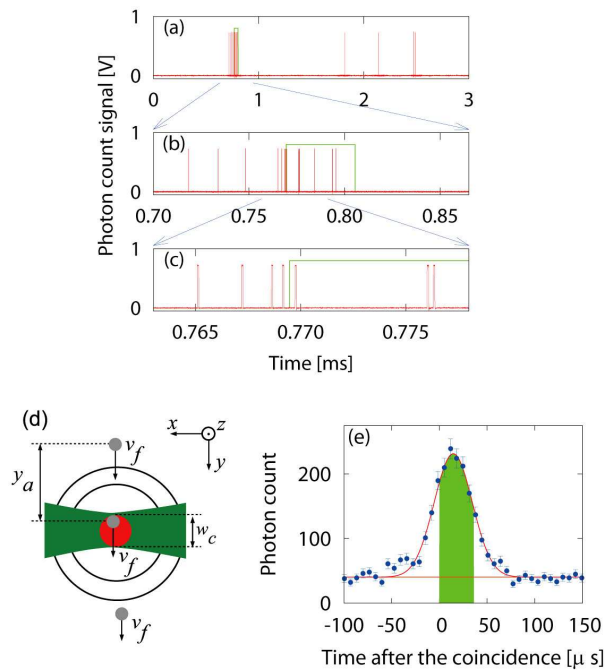}}\\
\caption{(Color online)
(a) Typical photon counting signals (red) and coincidence signals (green).
 (b), (c) Expanded views of the signal around the time when the spacing reaches $T_\mathrm{coin}$.
(d) The cavity mode viewed from the $z$-axis (red).
 The excitation beam is also shown (green).
(e) Time variation of the photon emission rate (blue dots).
The red curve is obtained by Gaussian fitting.
The green area indicates the time window of $T_\mathrm{win}$.
}
\label{nim}
\end{center}
\end{figure}

Typical photon counting signals obtained from SPCM2 are shown in Figs. \ref{nim}(a) through \ref{nim}(c).
The bunch of counts in Fig. \ref{nim}(a) represents an atom transit.
The average number of photon counts for a single atom transit is approximately 
10 count/atom.
This value is much larger than unity, in contrast to another method for single atom detection described in Ref. \cite{Terraciano09}.
The flux of atoms is approximately $F=1\times 10^{3}~\mathrm{s^{-1}}$, as shown in Fig. \ref{nim}(a), which is adjustable by reducing the loading time of the upper MOT.
A typical loading time is 0.3~s.
The mean spacing of the two neighboring atoms is estimated to be
 $y_a=v_f/F=0.3~\mathrm{mm}$,
 which is much larger than the beam waist size of the cavity, i.e., 19~$\mathrm{\mu m}$,
 as shown in Fig. \ref{nim}(d).
In other words, the average atom number in the cavity mode is much less than unity, $N_\mathrm{atom}=FT_\mathrm{transit}=0.1$.
Note that the photon count by the usual spontaneous emission from an atom is negligible under the present experimental condition.
The divergence angle of the cavity mode
 is $\theta=1\times10^{-2}$~rad. Therefore the detection efficiency becomes
 $q^\mathrm{(free)}=q\theta^2/4=8\times 10^{-6}$ and
the number of counts becomes
$q^\mathrm{(free)}\Gamma^\mathrm{(free)}(0)T_\mathrm{transit}=5\times 10^{-4}$
 for an atom transit,
 which is much smaller than the observed number of counts in each bunch of signals.
An enlarged view of the first bunch is shown in Fig. \ref{nim}(b).
The closer an atom approaches the mode axis,
 the narrower the spacing of the two photon counts becomes.
When the spacing decreases below $T_\mathrm{coin}=600~$ns,
 we judge that the position of the atom is close enough to the mode axis.
This threshold is valid because $T_\mathrm{coin}\sim (q \Gamma)^{-1}\ll w_c/v_f$ is satisfied,
 where $(q \Gamma)^{-1}$ is the time scale of the photon anti-bunching. 
We refer to this event as ``coincidence''.
After coincidence, the circuit outputs a logic pulse of duration $T_\mathrm{win}=36~\mathrm{\mu s}$, which we refer to as a ``coincidence signal''.
The timing and the duration of a coincidence signal generated by the photon counting signals are also shown in Figs. \ref{nim}(a) through \ref{nim} (c).

Figure \ref{nim}(e) shows a typical time variation of the photon emission rate,
 where the time is set to $t=0$ at the rising edge of the coincidence signal.
The observed signal can be fitted well by a Gaussian distribution.
The peak occurs $t_0=15~\mathrm{\mu s}$ after the coincidence,
 and the half width at $1/\sqrt{e}$ maximum is 20~$\mathrm{\mu s}$.
These findings indicate that the atom
 is well coupled to the cavity mode during $T_\mathrm{win}$.
We have confirmed that the signals during $T_\mathrm{win}$
 originate from single atoms
 by checking the anti-bunching of photons \cite{Hennrich05}.
The area of the offset from $t_1$ to $t_2$, say $S_\mathrm{offset}(t_1,t_2)$, is 0.6 times as large as the Gaussian area during the same time interval $S_\mathrm{gauss}(t_1,t_2)$, where we set $t_1=t_0-T_\mathrm{transit}/2$ and $t_2=t_0+T_\mathrm{transit}/2$.
On the contrary, the probability that more than one atom enter the cavity is 0.06 times smaller than the probability that one atom enters, where we assume that the atom number distributes according to a Poissonian distribution with the average number of $N_\mathrm{atom}$.
The atom number fluctuation cannot create such a large offset.
The offset can be understood as a results of accidental coincidences caused by ESEs of different atoms weakly coupled to the cavity. 
It does not affect the precision of the projective measurements but affects the success probability unless the ratio of $\eta(t_1,t_2)=S_\mathrm{gauss}/(S_\mathrm{gauss}+S_\mathrm{offset})$ keeps constant.
In our experimental conditions, $\eta(6~\mathrm{\mu s}, 36~\mathrm{\mu s}) =0.8$ .

\section{Projective measurements on a single nuclear spin}
\begin{figure}
\begin{center}
\scalebox{0.15}{\includegraphics{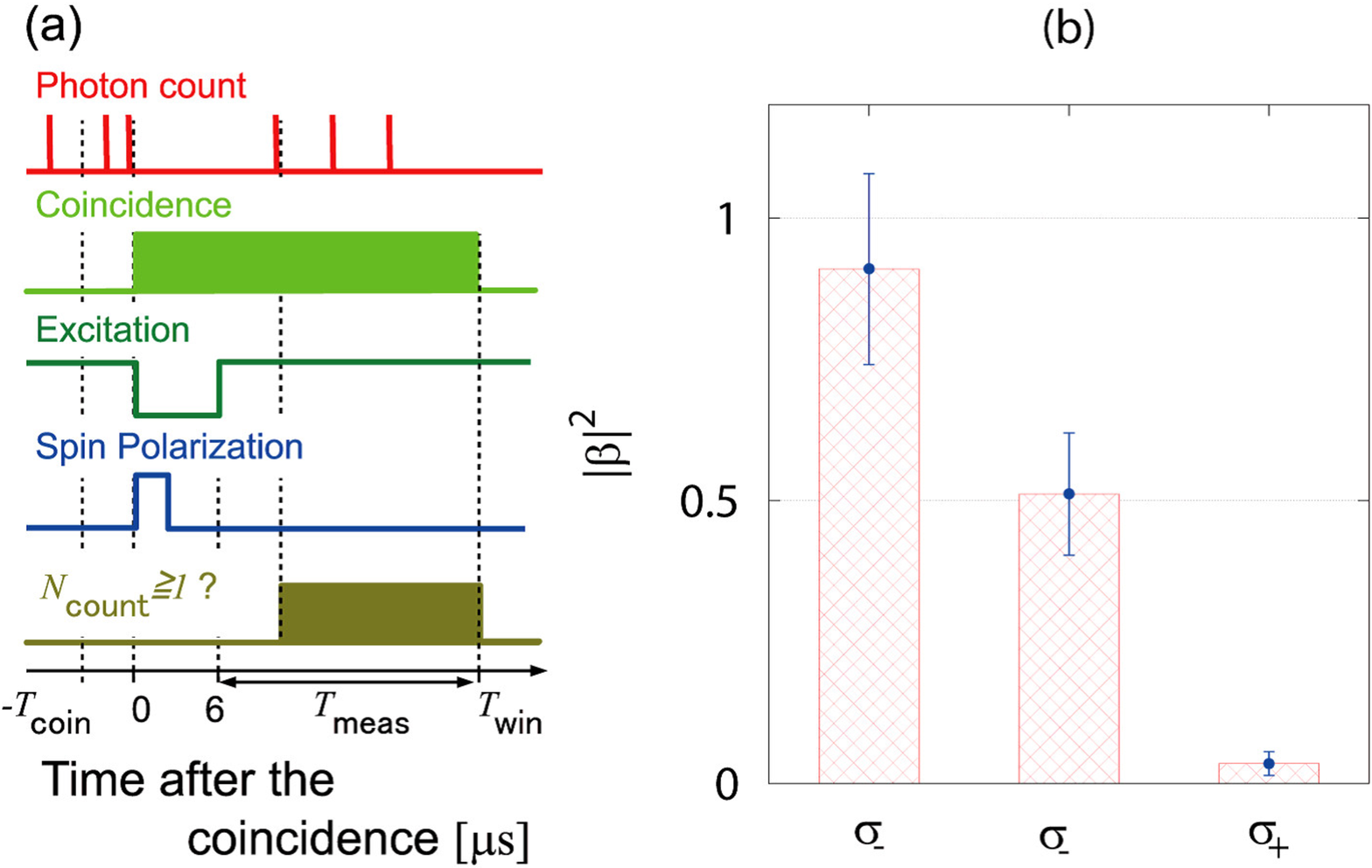}}
\caption{(Color online)
(a) Time chart for the projective measurement.
(b) Measurements of the diagonal elements $|\beta|^2=|\langle\downarrow\!|\psi\rangle|^2$ for the spin states prepared by
 $\{\sigma_-$, $\sigma_-^\perp$, $\sigma_+\}$ pulses.
}
\label{Fig4}
\end{center}
\end{figure}

Next, we demonstrate projective measurements on a single nuclear spin.
The procedure is shown in Fig. \ref{Fig4}(a) as a time chart.
After we confirmed that a $|\!\!\downarrow\rangle$ atom exists
 in the cavity mode by the coincidence,
 we turned off the excitation beam for $6~\mathrm{\mu s}$
 and injected a spin polarization pulse having a duration of $2~\mathrm{\mu s}$.
The alignments of the spin polarization pulses
 $\{\sigma_-$, $\sigma_-^\perp$, $\sigma_+\}$ are shown in Fig. \ref{setup5}.
The spin states prepared by $\{\sigma_-$, $\sigma_-^\perp$, $\sigma_+\}$ pulses
 are expected to be $\{|\!\!\downarrow\rangle$,
 $(|\!\!\downarrow\rangle+|\!\!\uparrow\rangle)/\sqrt{2}$,
 $|\!\!\uparrow\rangle\}$, respectively.
After initialization of the spin state,
 we turned on the excitation beam again and measured the number of counts
 $N_\mathrm{count}$ during $T_\mathrm{meas}=30~\mathrm{\mu s}$.
If the number of photon counts observed is greater than zero ($N_\mathrm{count}\ge 1$),
projection from an unknown spin state
 $|\psi\rangle$ to the $|\!\!\downarrow\rangle$ state is successful.
The measured value $N_\mathrm{count}$ for $|\psi\rangle=|\!\!\downarrow\rangle$ is $N_\mathrm{count}^{|\downarrow\rangle}=4$ on average.
Note that in the case of $N_\mathrm{count}=0$ the spin might be $|\!\!\uparrow\rangle$ due to the 
failure of the projective measurement.
Such a readout error probability is 0.02, estimated from a Poissonian statistics for a given average of $N_\mathrm{count}^{|\downarrow\rangle}$.
The diagonal element of the unknown spin state can be estimated
 by repeating the preparation of an unknown spin state.
When we prepare the $|\psi\rangle$ state $N_\mathrm{in}$ times,
 the number of successful projections becomes $N_\mathrm{suc}=\eta_0|\beta|^2 N_\mathrm{in}$, where $|\beta|^2=|\langle\downarrow\!|\psi\rangle|^2$, and
 $\eta_0$ corresponds to the success probability of the projective measurement for a coincidence.
The readout error of 0.02 is included in $\eta_0$.
Therefore, the $|\!\!\downarrow\rangle$ state can be automatically prepared by taking the coincidence condition without any spin polarization pulse at the success probability of $\eta_0$.
The values of $|\beta|^2$ initialized by $\{\sigma_-$, $\sigma_-^\perp$, $\sigma_+\}$ pulses are expected to be
$\{1$,$0.5$,$0\}$, respectively.
These values agree well with the experimental results, as shown in Fig. \ref{Fig4}(b) \cite{footnote1}.
The errors are caused by the statistical errors of $N_\mathrm{suc}$, the error of $\eta_0$ ($\eta_0=0.86\pm 0.09$), and the finite number of preparations, $N_\mathrm{in}=10^2$.
The measured value $\eta_0$ agrees well with the expectation discussed above.
The slight difference between the value for $\sigma_+$ initialization
 and the expected value ($|\beta|^2=0$) is probably caused
 by the imperfect circularity of the $\sigma_+$ light.
The readout error can be suppressed by just extending the measurement time with a moving lattice, for example.
If a single nuclear spin is observed during 60 $\mu s$, 
which is twice as large as these $T_\mathrm{meas}$,
$N_\mathrm{count}^{|\downarrow\rangle}$ will be 8 and the corresponding readout error can be less than $10^{-3}$.
This value satisfies a typical requirement in fault-tolerant quantum computing \cite{Myerson08}. 

\section{Systematic errors' evaluation}
\begin{figure}
\begin{center}
\scalebox{0.9}{\includegraphics{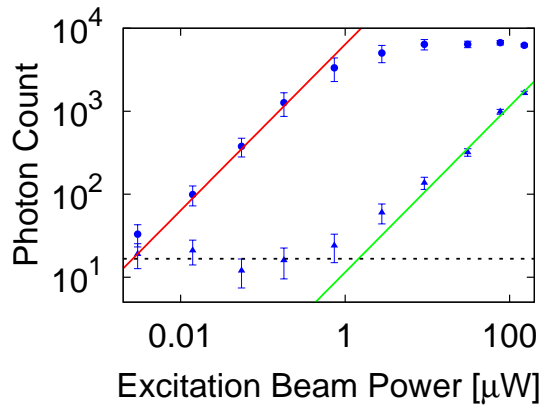}}
\caption{(Color online)
The numbers of measured photon counts for $\{\sigma_-'$, $\sigma_+'\}$ initialization 
 denoted as $\{$$N_{\sigma-'}$ (circle), $N_{\sigma+'}$ (triangle)$\}$.
$N_{\sigma-'}$ for the weak excitation beam and
$N_{\sigma+'}$ for the strong excitation beam
can be fitted by linear functions (red and green).
The dash-dotted line is the dark count level.
}
\label{Fig5}
\end{center}
\end{figure}

Finally, we indirectly evaluate the signal to noise ratio of these projective measurements using other experiment. After being released from the lower MOT, atoms are initialized by $\{\sigma_-'$, $\sigma_+'\}$ pulses, as shown in Fig. \ref{setup5}.
The number of counts during the entire measurement time $T_\mathrm{hold}$
 is accumulated without taking the coincidence condition.
Since atoms are exposed to the excitation beam longer than $T_\mathrm{meas}$, the probability of the spin flip increases compared to the procedure discribed by Fig. \ref{Fig4} (a). 
The lower limit of $S/N$ can be written as $S/N\ge N_{\sigma-'}/ N_{\sigma+'}$\,
 where $\{N_{\sigma-'}$, $N_{\sigma+'}\}$ is the number of counts for $\{\sigma_-'$, $\sigma_+'\}$ initialization.
We measured the dependence on the excitation beam power $P_\mathrm{total}$
of $\{N_{\sigma-'}$, $N_{\sigma+'}\}$, as shown in Fig. \ref{Fig5}.

We find that $N_{\sigma-'}/N_{\sigma+'}=4\times 10^2$ at $P_\mathrm{total}=0.9~\mathrm{\mu W}$, which is about one order of magnitude below the estimated $S/N$.
The difference is due to not only the spin flip during the extra excitation time, but also the noise floor of $N_{\sigma+'}$ limited by the dark count level.
The saturation of $N_{\sigma-'}$ at high excitation power is due to the saturation of the photon emission rate.
Therefore, the systematic error for the projective measurements is estimated as
 $\delta|\beta|^2\le(2+N_\mathrm{count}^{|\downarrow\rangle})N_{\sigma+'}/N_{\sigma-'}=2\times 10^{-2}$, well below the statistic errors. 
To the best of our knowledge, the state detections in electric spin cavity QED systems are typically performed with $T_\mathrm{meas}=100~\mathrm{\mu s}$, and $\delta|\beta|^2=0.52/30=2\times 10^{-2}$, for $2(F+1)=8$ degenerate hyperfine substates \cite{Boozer06}. Our system can address nondegenerate Zeeman substates, in other words a nuclear spin, with comparable performance.

\section{Summary}
In summary, we have constructed a cavity QED system with a diamagnetic atom of $\mathrm{^{171}Yb}$ and verified its nuclear spin state.
This can be used as a core technology for a large-scale one-way quantum computing.
Our result will expand the selection ranges of wavelength and transition strength available for atomic cavity QED experiments.
Especially, cavity QED systems with diamagnetic atoms will be interesting.
Diamagnetic atoms in a static electric field give the upper limit for the permanent electric dipole moment of atoms, the measurement of which can be used to test various theories beyond the Standard Model\cite{Griffith09, Dzuba07}.
The narrow transition of diamagnetic atoms leads to a new regime of cavity QED \cite{Meiser10}.
Diamagnetic atoms in optical lattice behave one of the most accurate clocks \cite{Takamoto05, Kohno09}.

We would like to thank T. Mukaiyama, T. Kishimoto, and S. Inouye for helpful discussions and their assistance in the experimental preparations.


\begin{thebibliography}{99}
\bibitem{footnote1} The result of $|\beta|^2\sim 0.5$ for $\sigma_-^\perp$ pulse does not provide a perfect evidence that we have generated $(|\!\!\downarrow\rangle+|\!\!\uparrow\rangle)/\sqrt{2}$. Measurements with a different basis are necessary for its verification.
\end{thebibliography}
\end{document}